\documentclass[twocolumn,superscriptaddress,amsmath,amssymb,aps,prx]{revtex4-2}

\usepackage{amsmath,amssymb}
\usepackage{graphicx}
\usepackage{dcolumn}
\usepackage{bm}
\usepackage{multirow}
\usepackage{textcomp}
\usepackage{url}
\usepackage{enumitem}
\usepackage{float}
\usepackage[dvipsnames]{xcolor}
\usepackage[colorlinks=true, allcolors=cyan]{hyperref}
\usepackage{orcidlink}
\newcommand{\bra}[1]{\langle{#1}|}
\newcommand{\ket}[1]{|{#1}\rangle}

\begin{document}

\title{Stochastic many-body perturbation theory for high-order calculations}

\author{X. Zhen\,\orcidlink{0009-0000-1806-4123}}
\affiliation{School of Physics, and State Key Laboratory of Nuclear Physics and Technology, Peking University, Beijing 100871, China}
\author{R. Z. Hu\,\orcidlink{0009-0002-8797-6622}}
\affiliation{School of Physics, and State Key Laboratory of Nuclear Physics and Technology, Peking University, Beijing 100871, China}
\author{J. C. Pei\,\orcidlink{0000-0002-9286-1304}}\email[]{peij@pku.edu.cn}
\affiliation{School of Physics, and State Key Laboratory of Nuclear Physics and Technology, Peking University, Beijing 100871, China}
\author{F. R. Xu\,\orcidlink{0000-0001-6699-0965}}\email[]{frxu@pku.edu.cn}
\affiliation{School of Physics, and State Key Laboratory of Nuclear Physics and Technology, Peking University, Beijing 100871, China}
\affiliation{Southern Center for Nuclear-Science Theory (SCNT), Institute of Modern Physics, Chinese Academy of Sciences, Huizhou 516000, China}

\date{\today}

\begin{abstract}
High-order perturbative \textit{ab initio} calculations are challenging due to the rapidly growing configuration space and the difficulty of assessing convergence.
In this letter, we introduce perturbation theory quantum Monte Carlo (PTQMC), a stochastic approach designed to compute high-order many-body perturbative corrections.
By representing the perturbative wave function with random walkers in configuration space, PTQMC avoids the exponential scaling inherent to conventional constructions of high-rank excitation operators. Benchmark calculations for the Richardson pairing model demonstrate that PTQMC accurately reproduces exact many-body perturbation theory (MBPT) coefficients up to 16th order, even in strongly divergent regimes. We further show that combining PTQMC with series resummation techniques yields stable and precise energy estimates in cases where the straightforward perturbative series fails. Finally, we propose the effective number of configurations, $e^{S}$, as a global measure of perturbative wave-function complexity that can be directly extracted within PTQMC. We demonstrate that the saturation behavior of $e^{S}$ provides a more reliable indicator of the validity of perturbative expansions than energy convergence alone.
\end{abstract}

\maketitle

\textit{Ab initio} calculations in nuclear physics are based on high-precision realistic nucleon--nucleon potentials, together with an approach to solve the many-body Schr\"odinger equation~\cite{machleidt2024recent, hergert2020guided, ekstrom2023ab, gandolfi2020atomic}. The mixing of possible configurations is one of the most common ways to construct correlated many-body wave functions~\cite{Thesis_CD, MBPT_CD, drischler2021chiral, drischler2020well, drischler2016neutron, tichai2020many, CC_Jiang, CC_Baardsen, CC_Hagen, IMSRG_Xu, IMSRG_Zhen, FCIQMC_Hu}. Many-body perturbation theory (MBPT) is a natural and standard extension of static perturbation theory to many-body systems~\cite{machleidt2024recent, hjorth2017advanced, shavitt2009many}. In MBPT, the coefficients of each configuration are determined perturbatively. One can obtain a higher-order many-body wave function through all lower orders, requiring summation over all configurations connected to the nonzero ones. Hence, high-order MBPT results can be acquired recursively, but only in rather small and simple systems~\cite{tichai2016hartree, hjorth2017advanced}. When it comes to realistic nuclear systems, the size of the configuration space is extremely large, making it impossible to obtain MBPT wave functions directly in configuration space. Most of these calculations are performed in operator space~\cite{RevModPhys.79.291, CC_Review, drischler2021chiral, IMSRG_Hergert}, instead of using matrix elements between configurations~\cite{FCIQMC_Hu}. Even so, high-order calculations exceeding fourth order have never been achieved in large systems~\cite{machleidt2024recent, MBPT_CD, Thesis_ZL}. 

In nuclear matter, fourth-order MBPT results are obtained using Monte Carlo methods to evaluate high-dimensional numerical integrals~\cite{MBPT_CD}. In the perturbation series, the total energy is expanded with respect to a coupling strength $\lambda$,
\begin{equation}
E = E_0 + \lambda E_1 + \lambda^2 E_2 + \cdots.
\end{equation}
Generally, this power series converges within a finite radius of convergence. When the strength of correlations in a system makes the radius of convergence smaller than $1$, the usual MBPT energy series evaluated at $E(\lambda=1)$ is likely to be asymptotic or even divergent~\cite{Thesis_ZL, leininger2000mo, kvaal2011computing}. Therefore, we cannot easily assess the difference between low-order MBPT energies and the exact ones. The only way to test the accuracy and uncertainty would be to go to higher orders and examine the convergence of the energy series systematically, which is nearly impossible in most cases. As a first illustration, we investigate high-order perturbative effects in the Richardson pairing model.

The Richardson pairing model is simple enough to be used for testing and benchmarking many-body calculations in nuclear physics~\cite{hjorth2017advanced, FCIQMC_Hu, brolli2025diagrammatic}. Our system contains four doubly degenerate energy levels and two pairs of particles, which makes the energy levels half-filled. The Hamiltonian is written as
\begin{equation}
\hat{H}=\delta \sum_{p=1}^{4} \sum_{\sigma=\uparrow,\downarrow} (p-1)a_{p\sigma}^\dagger a_{p\sigma} -\dfrac{g}{2} \sum_{p,q=1}^{4} a_{p\uparrow}^\dagger a_{p\downarrow}^\dagger a_{q\uparrow} a_{q\downarrow}.
\end{equation}
Without loss of generality, we set $\delta=1$. The coefficient $g$ determines the strength of correlations, with $g>0$ corresponding to attractive interactions and $g<0$ to repulsive ones. In this system, arbitrary-order MBPT energies can be calculated recursively, and their behaviors differ for different values of $g$.

\begin{figure}
    \centering
    \includegraphics[width=1.00\linewidth]{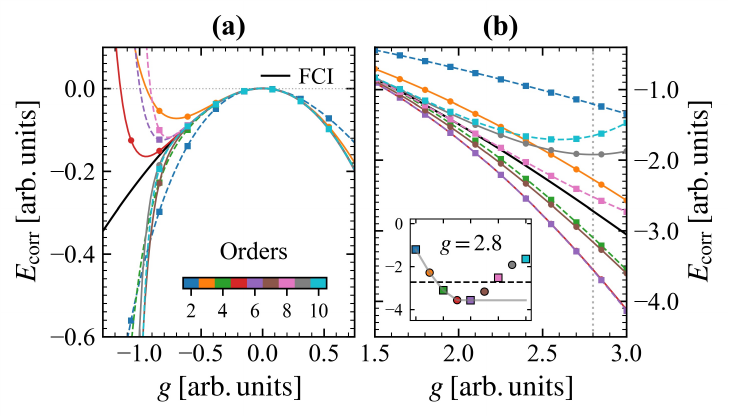}
    \caption{The summed correlation energies from high-order MBPT calculations for Richardson pairing model, in arbitrary units. Dashed lines with square markers denote even orders, while solid lines with circular markers denote odd orders. The solid black line indicates the exact FCI value. (a) The area corresponding to $-1.3<g<0.75$. (b) The area corresponding to $1.5<g<3.0$. In the inset, the order-by-order correlation energies at $g = 2.8$ are shown to emphasize the pseudoconvergence behavior.}
    \label{fig:first}
\end{figure}

In Fig.~\ref{fig:first}, the summed correlation energies from high-order MBPT calculations up to tenth order are shown. In panel~(a), we find that the MBPT results converge well when $-0.5 \lesssim g \lesssim 0.75$. In the repulsive region, the MBPT calculations become strongly divergent as the absolute value of $g$ increases. In this regime, the naive perturbative calculations are no longer reliable. These results are consistent with well-known findings~\cite{faribault2011gaudin, snyman2006analysis, claeys2015eigenvalue}. In panel~(b), some new features of MBPT arise. In previous perturbative calculations, it is expensive to use next-order calculations to evaluate how far we are from the ultimate converged value. Therefore, a practical method is to use the order-by-order ratio. If the highest order we can afford is MBPT(4), then if
\begin{equation}
    \Bigg|\frac{\Delta E_4}{\Delta E_3}\Bigg| < \Bigg|\frac{\Delta E_3}{\Delta E_2}\Bigg| < \Bigg|\frac{\Delta E_2}{\Delta E_1}\Bigg| < 1,
    \label{eq:ratio}
\end{equation}
the MBPT(4) value is usually regarded as a good approximation of the final converged value. We observe a counterexample when $g>1.5$ in the Richardson model. In panel~(b), the convergence pattern seems to be satisfied from MBPT(2) to MBPT(6), with the ratio like in Eq.~\eqref{eq:ratio} decreasing order by order. The ratio $|\Delta E_6/\Delta E_5|$ is almost zero, and it seems that we can sufficiently believe that the MBPT(6) energy is very close to the full configuration-interaction (FCI) result. However, if one continues to calculate the seventh-order correction, the situation becomes worse, as the ratio increases again. In fact, the perturbative series in this region is not convergent, but appears to be well disguised. The MBPT(6) energy, although it seems to be converged, is in fact far from the FCI result. This divergence pattern is obviously different from the strongly oscillating ones and can be misleading to get a reliable approximation. These findings imply that in specific systems, for example in strongly correlated situations in nuclear matter, it is still necessary to use higher order-by-order results to analyze the accuracy and uncertainty of MBPT calculations.

To perform higher-order calculations in operator space, one needs to identify all connected diagrams and sum over all possible particle--hole lines in the corresponding Hugenholtz diagrams~\cite{shavitt2009many}. The number of diagrams grows rapidly with increasing order, resulting in a computational cost at the $n$-th order that scales at least as $O(N^{2n})$, where $N$ is the number of single-particle states. For typical calculations with $N \sim 10^3$, the cost of high-order calculations becomes enormous and impractical. Moreover, if residual three-nucleon interactions are included, the number and topology of the diagrams become even more complicated and are computationally prohibitive. On the other hand, as mentioned in the previous section, recursively calculating high-order perturbative contributions also faces severe difficulties. The core challenge is how to identify the most important configurations among all connected structures. Inspired by the efficient algorithm of full configuration-interaction quantum Monte Carlo (FCIQMC)~\cite{booth2009fermion, shepherd2012full, FCIQMC_Hu, booth2013towards}, we have developed a perturbative variant, referred to as the perturbation theory quantum Monte Carlo~(PTQMC). The capitalized abbreviation PTQMC highlights its connection to MBPT and distinguishes it from projection-type Monte Carlo approaches sometimes abbreviated as ptQMC~\cite{pt1, pt2}. 

The PTQMC algorithm starts from a simple observation: the $n$th-order MBPT energy correction can be viewed as the summation over all possible routes in configuration space, both beginning and ending at the reference state,
\begin{equation}
    \ket{0} \to \ket{I_1} \to \cdots \to \ket{I_{n-1}} \to \ket{0}.
\end{equation}

\begin{figure*}[htbp]
    \centering
    \includegraphics[width=0.75\linewidth]{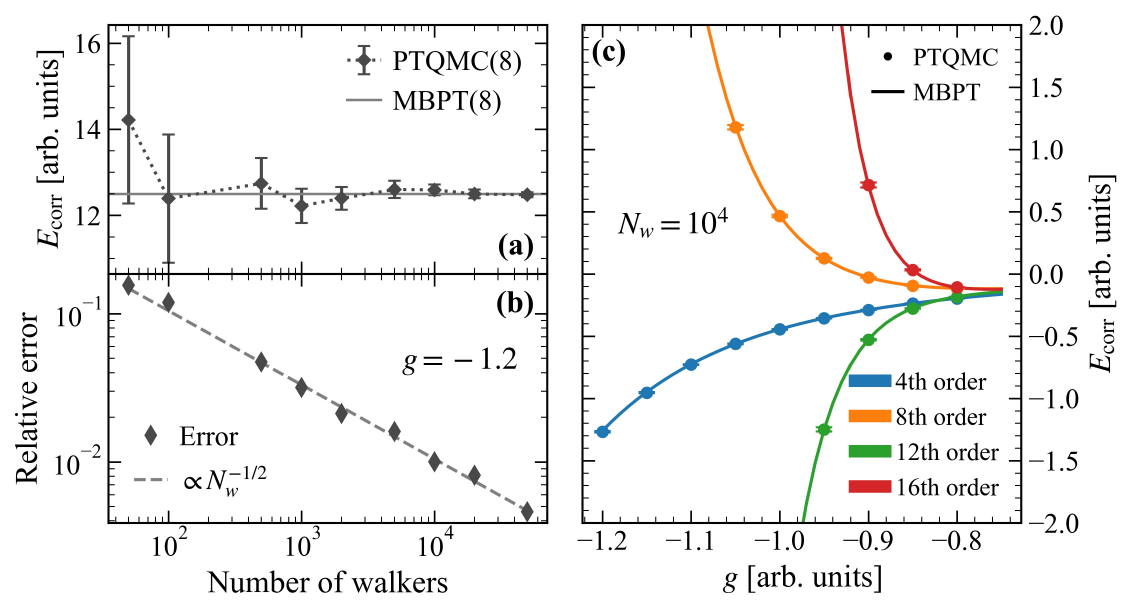}
    \caption{Benchmarking PTQMC against exact MBPT calculations. (a) Eighth-order PTQMC correlation energy $E_{\mathrm{corr}}$ for the Richardson pairing model at $g=-1.2$, compared with the exact MBPT result. (b) Relative error of the PTQMC energy with respect to the exact MBPT value as a function of the number of walkers, demonstrating the expected statistical scaling $\propto N_\mathrm{w}^{-1/2}$. (c) PTQMC correlation energies computed up to 16th order over the coupling range $-1.2 < g < -0.76$, a regime where the conventional MBPT series is strongly divergent.}
    \label{fig:second}
\end{figure*}
Here $I$ denotes a single configuration, and the ``$\to$'' symbol indicates that two configurations are connected via the perturbation, $\bra{I}\hat{H}_1\ket{J} \neq 0$. In the summation over all possible routes, not all paths contribute equally to the final energy correction. Our goal is to use a stochastic approach to identify the most important contributions at each order, as well as their correct cumulative effects toward higher orders.

As a perturbative method, we decompose the many-body Hamiltonian into an exactly solvable part and a residual part,
\begin{equation}
    \hat{H} = \hat{H}_0 + \hat{H}_1,
\end{equation}
where $\hat{H}_0$ is chosen to be diagonal in the many-body basis and exactly solvable, and $\hat{H}_1$ is treated as a perturbation. Similar to FCIQMC, the perturbative wave function is represented stochastically by random walkers, which are distributed over many-body configurations.
The average walker population on a configuration is proportional to its perturbative coefficient at a given order.
The $n$th-order MBPT wave function is written as~\cite{shavitt2009many}
\begin{equation}
    \ket{\Psi^{(n)}} = \sum_I c_I^{(n)} \ket{\Phi_I},
\end{equation}
where $\ket{\Phi_I}$ denotes a many-body basis state and $c_I^{(n)}$ is its perturbative coefficient.
The corresponding signed walker population is denoted by $w_I^{(n)}$.
The walker weights satisfy the standard recursive relation of Rayleigh--Schr\"odinger many-body perturbation theory~\cite{shavitt2009many, tichai2020many},
\begin{align}
    w_I^{(n+1)} &=
    \frac{1}{\Delta_{I0}}\left[
    \sum_J \bra{\Phi_I}\hat{H}_1\ket{\Phi_J} \, w_J^{(n)}\right.
    \label{eq:iter} \\
    &\left.\quad - \sum_{m=1}^{n} E^{(m)} w_I^{(n-m+1)}\right], \nonumber
\end{align}
with the initial conditions
\begin{equation}
    w_I^{(0)} = \delta_{I0}, \qquad
    w_I^{(1)} = \frac{\bra{\Phi_I}\hat{H}_1\ket{\Phi_0}}{\Delta_{I0}}.
\end{equation}
Here, the reference state $\ket{\Phi_0}$ is the ground-state eigenvector of $\hat{H}_0$, and the M\o ller-Plesset energy denominator is defined as~\cite{shavitt2009many}
\begin{equation}
    \Delta_{I0} \equiv E_0^{(0)} - E_I^{(0)}
    = \bra{\Phi_0}\hat{H}_0\ket{\Phi_0} - \bra{\Phi_I}\hat{H}_0\ket{\Phi_I}.
\end{equation}

Equation~\eqref{eq:iter} is formally identical to the deterministic MBPT recursion relation and will be evaluated stochastically.
Starting from the first-order distribution $w_I^{(1)}$, walkers are propagated order by order.
For each walker residing on configuration $\ket{\Phi_J}$, spawning events to connected configurations $\ket{\Phi_I}$ are proposed stochastically.
The magnitude of the spawning probability is proportional to
$|\bra{\Phi_I}\hat{H}_1\ket{\Phi_J}| / |\Delta_{I0}|$,
while the sign of the spawned walker is determined by the sign of
$\bra{\Phi_I}\hat{H}_1\ket{\Phi_J}w_J / \Delta_{I0}$.
In this way, the stochastic process provides an unbiased estimator of the first term in Eq.~\eqref{eq:iter}, which costs the most in usual computations.

After all spawning events at a given order are completed, all signed walkers residing on the same configuration are collected and summed yielding the updated weight distribution $w_I^{(n+1)}$. The deterministic correction terms proportional to the lower-order energy contributions in Eq.~\eqref{eq:iter} are applied simultaneously. After each iteration, the $(n+1)$th-order correlation energy can be estimated as
\begin{equation}
    E^{(n+1)} = \sum_I \bra{\Phi_0}\hat{H}_1\ket{\Phi_I} \, w_I^{(n)},
\end{equation}
which corresponds to the standard MBPT energy correction. Unlike projector quantum Monte Carlo methods~\cite{wlazlowski2014auxiliary, lynn2016chiral, roggero2014quantum, arthuis2023quantum, carlson2015quantum, FCIQMC_Hu}, the present approach does not evolve the wave function toward the exact ground state. Instead, each perturbative order is sampled independently according to the exact MBPT recursion relations, allowing direct access to the order-resolved structure of the many-body wave function. More details of the algorithm can be found in the Supplemental Material~\cite{SM}.

It is worth emphasizing that the PTQMC algorithm relies solely on the availability of Hamiltonian matrix elements between many-body configurations, without invoking any additional model-specific structure~\cite{FCIQMC_Hu}. When the Møller--Plesset perturbation series is convergent, the corresponding perturbative expansion uniquely represents the exact many-body wave function. PTQMC achieves this by sampling perturbative orders rather than the full configuration space, which can lead to a substantial reduction in computational cost in weakly to moderately correlated systems.
Although the present work focuses on the Richardson pairing Hamiltonian as a well-controlled benchmark, the above considerations indicate that the applicability of PTQMC is not restricted to pairing models and can be extended to general nuclear many-body Hamiltonians.

\begin{figure*}[htbp]
    \centering
    \includegraphics[width=0.7\linewidth]{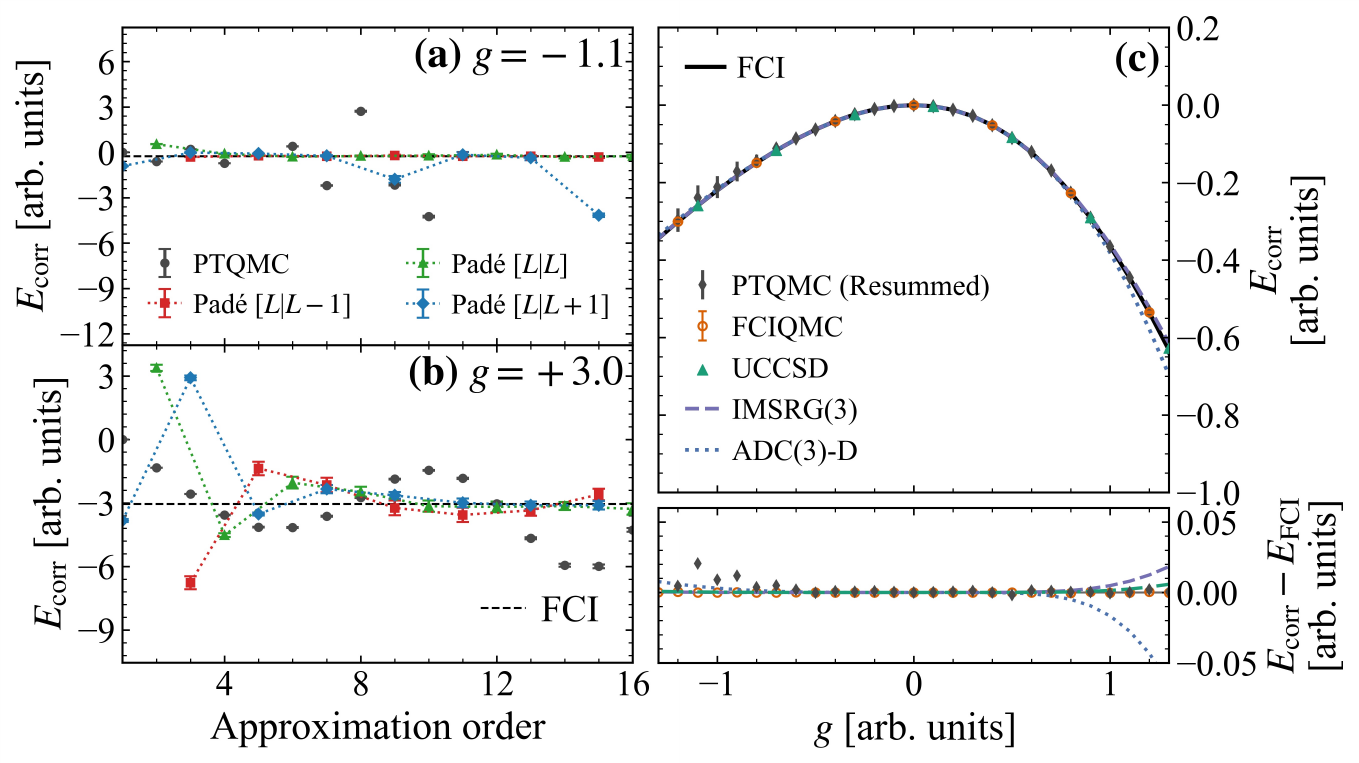}
    \caption{Resummation of high-order PTQMC data using Pad\'e approximation and comparison with exact value. (a) Correlation energy as a function of approximation order at $g=-1.1$. (b) Same as (a) but for $g=+3.0$. (c) Correlation energy as a function of coupling strength $g$ obtained from the resummed PTQMC results, compared with different kinds of nonperturbative calculations. The lower panel shows the deviation of each method from the exact FCI result.}
    \label{fig:third}
\end{figure*}
As a newly developed many-body approach, it is essential to benchmark PTQMC against exact results available. Figure~\ref{fig:second} presents a detailed comparison between PTQMC and exact MBPT calculations for the Richardson pairing model. In Fig.~\ref{fig:second}(a), we show the eighth-order correlation energy at $g=-1.2$. The PTQMC(8) results are represented by dark-gray diamonds with statistical error bars, while the exact MBPT(8) value is shown as a light-gray horizontal line. The PTQMC estimates are statistically consistent with the exact MBPT value, and the statistical uncertainty decreases systematically with increasing walker number. Figure~\ref{fig:second}(b) displays the relative error of the PTQMC energy with respect to the exact MBPT result as a function of the number of walkers. The observed $N_\mathrm{w}^{-1/2}$ scaling, shown by the light-gray dashed line, confirms the expected statistical convergence behavior of the stochastic Monte Carlo sampling. In Fig.~\ref{fig:second}(c), we fix a sufficiently large walker number $N_\mathrm{w} = 10^4$ and benchmark PTQMC (dots) against exact MBPT results (solid lines) over a strongly divergent coupling regime at different high orders, indicated by different colors. In this regime, the high-order MBPT contributions exhibit large oscillations and do not converge as a series. Nevertheless, PTQMC provides a statistically stable and unbiased stochastic evaluation of the corresponding high-order MBPT coefficients, accurately reproducing the exact perturbative results up to 16th order. These benchmarks demonstrate that PTQMC faithfully reproduces exact perturbative coefficients where available, even in regimes where the conventional MBPT series exhibits strong oscillatory behavior.

Importantly, the breakdown of convergence in high-order perturbation theory does not imply that the corresponding perturbative coefficients lack physical significance. Even when the series exhibits strong oscillations, these high-order contributions often contain essential information about the underlying many-body correlations. A key motivation for accessing such high-order data is their use in resummation schemes, through which stable physical predictions can be obtained at a substantially reduced computational cost.

Figure~\ref{fig:third} illustrates the resummation of high-order PTQMC perturbative results, including statistical uncertainties, and demonstrates how physically meaningful information can be extracted from a strongly divergent series. All PTQMC calculations are conducted using $N_\mathrm{w} = 10^4$. As a simple and widely used resummation scheme, the Pad\'e approximation was employed~\cite{pade1892representation, pade_book}, with statistical uncertainties propagated from the PTQMC input data. The Pad\'e approximation represents a perturbative series by a ratio of two polynomials and is known to capture nonanalytic structures such as singularities.
It has therefore been used to extend the applicability of MBPT energies in existing researches~\cite{roth2010pade, sammarruca2018nuclear, langhammer2012spectra, Thesis_ZL}. In Figs.~\ref{fig:third}(a) and \ref{fig:third}(b), we show Pad\'e approximants of order $[L/M]$, with $M=L, L\pm1$, indicated by different colored dotted lines.
The raw PTQMC perturbative results are shown as gray circles, while the exact FCI energy is indicated by the black dashed line. The approximation order corresponds to $L+M$, or equivalently PTQMC$(n)$. As shown in Fig.~\ref{fig:third}(a), for $g=-1.1$, the high-order perturbative contributions exhibit strong oscillations, rendering the bare perturbative sequence itself unreliable as a predictor of the physical energy. In contrast, the Pad\'e resummation yields a stabilized estimate that accurately reproduces the exact result once sufficient high-order information is included. In Fig.~\ref{fig:third}(b), corresponding to $g=+3.0$, where the perturbative series displays quasiperiodic oscillatory behavior, the Pad\'e approximants converge more gradually. Nevertheless, once the resummed values stabilize, they provide physically meaningful energy estimates, in contrast to the strongly oscillatory raw perturbative results, thereby avoiding spurious apparent convergence at low perturbative orders.

In Fig.~\ref{fig:third}(c), we compare the resummed PTQMC results with several nonperturbative many-body approaches~\cite{Heinz2020IMSRG3B, marino2024diagrammatic, FCIQMC_Hu, bartlett1988expectation} over the coupling range $-1.3 \le g \le 1.3$.
Different methods are indicated by distinct marker styles, and the lower panel shows their deviations from the exact FCI energies. The resummed PTQMC results are obtained from stable Pad\'e approximants, without restricting to a fixed choice of $L$ and $M$. It is found that low-order nonperturbative methods, such as ADC(2) and IMSRG(2), fail to provide reliable correlation energies over the entire coupling range here~\cite{brolli2025diagrammatic, hjorth2017advanced}. In contrast, the resummed high-order PTQMC results exhibit very good agreement with the exact FCI energies across the full interval, and even outperform several higher-order nonperturbative approaches in the strong-coupling region $g \gtrsim 1.0$. These results highlight the effectiveness of resumming high-order PTQMC perturbative data. Once reliable high-order information is available, physically meaningful energies can be obtained with a comparatively simple stochastic algorithm. The computational cost of PTQMC scales approximately as $\mathcal{O}(n N_\mathrm{w})$, where $n$ is the number of independent Monte Carlo runs and $N_\mathrm{w}$ is the walker population. On the other hand, when the absolute values of the high-order MBPT contributions become very large, a larger walker number is required to control statistical uncertainties, leading to increased computational cost~\cite{FCIQMC_Hu}. Compared to conventional nonperturbative methods, this introduces a trade-off between stochastic sampling effort and deterministic computational expense. Assessing the optimal balance between these approaches is therefore essential for practical applications of high-order PTQMC calculations of nuclear matter in future.

\begin{figure}
    \centering
    \hspace*{-1cm}
    \includegraphics[width=0.75\linewidth]{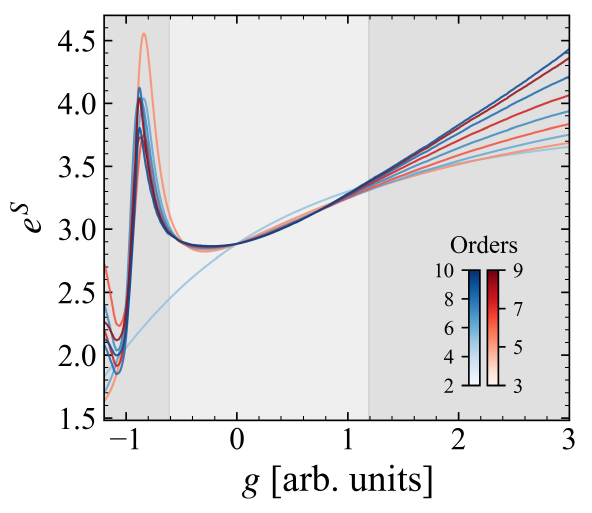}
    \caption{Effective number of configurations $e^{S}$ as a function of the coupling strength $g$ for different perturbative orders. The dark-shaded regions mark coupling intervals where $e^{S}$ exhibits divergence.}
    \label{fig:fourth}
\end{figure}

Finally, we comment on the structure of the many-body wave function at high perturbative orders. In conventional MBPT calculations, obtaining an explicit approximation to the wave function requires substantial additional effort~\cite{shavitt2009many, hjorth2017advanced}, involving the evaluation of a rapidly increasing number of Hugenholtz diagrams. In contrast, PTQMC samples many-body configurations directly, which naturally enables direct access to the perturbative wave-function structure. Because the detailed distribution of configuration weights becomes increasingly complex at high orders, it is impractical to analyze the wave function on a configuration-by-configuration basis.
Instead, we introduce a global measure to characterize the overall complexity of the perturbative wave function.
In quantum information theory, the Shannon entropy~\cite{shannon1948mathematical} constructed from the probability of finding a system in a given configuration, has been widely used to quantify wave-function complexity~\cite{sokolov1998invariant, flambaum2001entropy, armstrong2012complexity, kota1998information}. Within the PTQMC framework, the Shannon entropy at perturbative order $n$ is defined as
\begin{equation}
    S^{(n)} = - \sum_I p_I^{(n)} \ln p_I^{(n)},
\end{equation}
where $p_I^{(n)} = |w_I^{(n)}|^2 / \sum_J |w_J^{(n)}|^2$ is the normalized probability of configuration $\ket{\Phi_I}$ at order $n$. The exponential of the entropy, $e^{S^{(n)}}$, can be interpreted as the effective number of configurations contributing to the perturbative wave function and thus provides an intuitive measure of wave-function complexity~\cite{armstrong2012complexity, kota1998information}. 

Figure~\ref{fig:fourth} shows the effective number of configurations, $e^{S}$, as a function of the coupling strength $g$ for different perturbative orders.
The dark-shaded regions indicate coupling intervals where $e^{S}$ exhibits poor convergence up to tenth order, quantified by the criterion
\begin{equation}
    \Bigg|\frac{e^{S^{(10)}} - e^{S^{(9)}}}{e^{S^{(9)}}}\Bigg| \ge 0.2\%.
\end{equation}
The numerical threshold is introduced as a practical diagnostic and is not tied to any specific model property.
We have verified that varying this threshold within a reasonable range
(e.g., $0.1\%$--$0.5\%$) leads to the same qualitative identification of
poorly convergent regions. These regions clearly coincide with coupling ranges where the perturbative energy series displays divergent or strongly oscillatory behavior.
Such behavior reflects a continuous spreading of the wave-function weight over an increasing number of configurations, indicating that the complexity of the perturbative wave function has not stabilized with increasing perturbative order.
In contrast, outside the divergent regime, the effective number of configurations exhibits a much weaker dependence on the perturbative order and rapidly approaches saturation.
In these regions, the structure of the perturbative wave function becomes stable, suggesting that the associated high-order perturbative information is well behaved and amenable to resummation.

These observations indicate that the saturation behavior of $e^{S}$ provides a practical and physically motivated criterion for assessing the reliability of resummation based on high-order perturbative data.
When the effective number of configurations remains bounded as the perturbative order increases, resummation procedures can be expected to yield stable and physically meaningful results.
Conversely, a rapid growth of $e^{S}$ signals a breakdown of such approaches due to uncontrolled wave-function complexity. Because inaccurate or poorly converged wave functions may nevertheless reproduce seemingly well-converged energies, relying solely on energy convergence can lead to spurious apparent convergence, as illustrated in Fig.~\ref{fig:first}. In contrast, when $e^{S}$ itself shows clear saturation, low-order perturbative results are unlikely to suffer from such pseudoconvergence and can therefore be regarded as reliable.

In this work, we have introduced perturbation theory quantum Monte Carlo (PTQMC), a stochastic approach that enables the systematic evaluation of high-order many-body perturbative contributions in nuclear systems. By reformulating Rayleigh–Schrödinger many-body perturbation theory as a stochastic sampling problem in configuration space, PTQMC bypasses the explicit construction of high-order diagrams and avoids the prohibitive scaling of conventional deterministic implementations. Using the Richardson pairing model as a benchmark, we demonstrated that PTQMC faithfully reproduces exact perturbative coefficients up to high orders, including regimes where the perturbative series is strongly divergent or exhibits deceptive apparent convergence at intermediate orders. These results highlight that the failure of low-order MBPT cannot be reliably diagnosed without explicit access to higher-order information. 

When combined with Padé resummation, the high-order PTQMC data yield stable and accurate ground-state energies over a wide coupling range and, in certain strong-correlation regimes, outperform several commonly used high-order nonperturbative many-body methods. A central advantage of PTQMC is its direct access to the perturbative wave-function structure. We showed that the complexity of the perturbative wave function, quantified by the effective number of configurations derived from the Shannon entropy, provides crucial insight beyond energy convergence alone. In particular, the saturation of this quantity with increasing perturbative order signals well-behaved perturbative information amenable to resummation, while its continuous growth indicates uncontrolled wave-function spreading and unreliable perturbative descriptions. This observation explains why seemingly converged low-order energies can still be far from the exact result and provides a physically motivated criterion for identifying spurious convergence. Furthermore, reliable access to high-order perturbative coefficients may also provide insights into the analytic structure of many-body expansions and emergent phenomena in strongly correlated systems. 

The PTQMC framework offers a flexible and scalable route to explore high-order perturbative physics in nuclear many-body systems. Its stochastic nature makes it particularly appealing for applications where traditional diagrammatic or operator-based approaches become impractical, including systems with strong correlations and residual three-nucleon interactions. The present results lay the groundwork for future applications of PTQMC to realistic nuclear matter and finite nuclei, as well as for the systematic assessment of perturbative uncertainties in \textit{ab initio} nuclear theory.

The authors are grateful for communications with M. Heinz on IMSRG. We would also like to thank H.~Y.~Shang and C.~J.~Jiang for other useful discussions. This work is supported by the National Key R\&D Program of China under Grant No. 2023YFE0101500, NO.2023YFA1606400 and No.2024YFA1610900; the National Natural Science Foundation of China under Grants No.12335007, No.12535008 and No.12475118. We acknowledge the High-Performance Computing Platform of Peking University for providing computational resources.

\textit{Data Availability.} The data that support the findings of this study are available in GitHub Repository~\cite{data_git}.

\bibliographystyle{modified-apsrev4-2}
\bibliography{reference}

\end{document}


\title{Supplemental Material for: Stochastic many-body perturbation theory for high-order calculations}

\author{X. Zhen\,\orcidlink{0009-0000-1806-4123}}
\affiliation{School of Physics, and State Key Laboratory of Nuclear Physics and Technology, Peking University, Beijing 100871, China}
\author{R. Z. Hu\,\orcidlink{0009-0002-8797-6622}}
\affiliation{School of Physics, and State Key Laboratory of Nuclear Physics and Technology, Peking University, Beijing 100871, China}
\author{J. C. Pei\,\orcidlink{0000-0002-9286-1304}}\email[]{peij@pku.edu.cn}
\affiliation{School of Physics, and State Key Laboratory of Nuclear Physics and Technology, Peking University, Beijing 100871, China}
\author{F. R. Xu\,\orcidlink{0000-0001-6699-0965}}\email[]{frxu@pku.edu.cn}
\affiliation{School of Physics, and State Key Laboratory of Nuclear Physics and Technology, Peking University, Beijing 100871, China}
\affiliation{Southern Center for Nuclear-Science Theory (SCNT), Institute of Modern Physics, Chinese Academy of Sciences, Huizhou 516000, China}

\maketitle
\section{MBPT in configuration space}
The MBPT correlation energies in configuration space, as defined in Eqs.~(7)–(9) of the main text, depend solely on the weights of configurations that are connected to the reference state $\Phi_0$~\cite{shavitt2009many, tichai2020many}. We denote by $\Phi[n]$ the set of configurations that are connected to $\Phi_0$ by at least $n$ successive applications of the Hamiltonian operator. Contributions from configurations with $\Phi[n>1]$ are fully encoded in the weights of $\Phi[1]$ configurations and first appear in the perturbative expansion at order $2n$. If the MBPT series is well converged, these contributions decrease with increasing $n$. In other words, the importance of configurations that are farther away from the reference state $\Phi_0$ is expected to be progressively suppressed.

In our context, two configurations are defined to be connected by the perturbation $\hat{H}_1$ if
$\langle \Phi_1 | \hat{H}_1 | \Phi_2 \rangle \neq 0$.
In nuclear calculations, $\hat{H}_1$ may be chosen to include both normal-ordered two-nucleon ($N\!N$) and three-nucleon ($3N$) contributions, such that the corresponding matrix elements can be evaluated using the standard Slater--Condon rules~\cite{PhysRev.34.1293, PhysRev.36.1121}.
When formulated in operator-space language, all possible selections of single-particle indices must be enumerated, leading to a rapidly increasing combinatorial complexity and a large number of Hugenholtz diagrams~\cite{arthuis2019adg}.
In configuration space, by contrast, the structure of the perturbative contributions becomes significantly simpler and more transparent: the only required ingredient is to identify, in a stochastic and unbiased manner, all configurations that are connected to a given configuration.
This makes configuration-space MBPT particularly well suited for stochastic sampling and random-walk strategies, and potentially the only viable route toward accessing high perturbative orders. As an illustrative example, when three-nucleon indices are included, the number of connected Hugenholtz diagrams contributing to the correlation energy grows rapidly with perturbative order and reaches ${\cal O}(10^4)$ by fifth order (MBPT5). 
In contrast, when formulated in configuration space, the correlation energy at each order can be represented by a single topological class of configuration-space paths, with the increasing complexity absorbed into the stochastic propagation of configuration weights. Building on experience gained from the development of FCIQMC method for nuclear matter, we are well positioned to implement and exploit such strategies.
The concrete algorithm is described in the next section.

\section{Details of PTQMC algorithm}

In this section, we provide additional details of the PTQMC algorithm.
The overall structure of the algorithm, as well as the associated uncertainty analysis, closely follows that developed in our previous work on FCIQMC.
We therefore focus on the essential differences specific to PTQMC, highlighting the steps that depart from standard FCIQMC implementations and clarifying which ingredients are particular to the pairing model and which are not.
This discussion is intended to demonstrate that the present framework is readily extensible to nuclear matter and other realistic many-body systems.
We do not repeat implementation details that are identical to our previous FCIQMC work, such as the excitation-generation algorithm used to generate connected determinants. A complete description can be found in~\cite{FCIQMC_Hu}.

In PTQMC, each walker $a$ resides on a Slater determinant $\ket{\Phi}$ in configuration space and carries a complex phase , denoted by a complex unit vector $\hat{n}_a$.
The sum of all walkers on a given configuration $\ket{\Phi_I}$ defines the complex population $\hat{N}_I$, which is proportional to the coefficient $C_I$ associated with that configuration, such that $|C_I|^2$ represents its contribution to the many-body wave function $\ket\Psi$.
In contrast to FCIQMC, PTQMC does not aim to solve an imaginary-time master equation~\cite{FCIQMC_Hu}, nor does it involve the integration of a continuous differential equation.
Instead, each stochastic update corresponds to a discrete propagation in configuration space, extending the route through successively connected configurations.
In this formulation, the length of a route directly reflects the perturbative order: each additional propagation step increases the order of the many-body perturbation expansion.
A PTQMC calculation consists of three steps:

1. \textit{Initialization.}
We initialize the walker population to represent the first-order amplitudes $w^{(1)}$ (which enter the second-order energy estimator $E^{(2)}$ in Eq.~(8) of the main text).
To control the population size, we monitor the total walker number
\begin{equation}
  N_w = \sum_{I}\left|\hat{N}_I\right|,
\end{equation}
and normalize it to a target value $N_0$.
Because both the walker population and the number of occupied configurations can grow rapidly with perturbative order, we apply a renormalization step after each order update by rescaling all populations such that $N_w \rightarrow N_0$.

2. \textit{Spawning.}
For each walker on $\ket{\Phi_J}$, we randomly choose a connected determinant $\ket{\Phi_I}$ with proposal probability $p_\mathrm{gen}(I|J)$ using the excitation-generation algorithm~\cite{FCIQMC_Hu}.
A spawn from $J$ to $I$ is attempted with probability
    \begin{equation}
        p_\mathrm{s}(I|J) = \dfrac{1}{\Delta_{I0}}\dfrac{|H_{IJ}|}{p_\mathrm{gen}(I|J)},
    \end{equation}
and the phase of the spawned walker on $\ket{\Phi_I}$ is given by $-\hat{h}_{IJ}\hat{n}_J$, where $\hat{h}_{IJ}=H_{IJ}/|H_{IJ}|$.

The stochastic spawning corresponds to the first term in Eq.~(7) of the main text.
The remaining terms in Eq.~(7), which enforce the cancellation of disconnected diagrams, are evaluated deterministically from previously accumulated lower-order energies and amplitudes. This requires only simple multiplications and therefore incurs negligible additional cost, provided these quantities are stored at each order.

3. \textit{Annihilation and evaluation.}
After spawning, all contributions on the same determinant $\ket{\Phi_I}$ are combined (annihilated) by summing their phases, yielding the updated complex population
\begin{equation}
  \hat{N}_I = \sum_{a\in\Phi_I}\hat{n}_a = N_I \hat{n}_I.
\end{equation}
Once the redistribution is completed, the MBPT correlation energy is evaluated using Eq.~(10) of the main text.
A normalization factor has to be included to recover the unbiased estimator of the perturbative contribution at the corresponding order.

For the pairing model benchmarks considered here, the walker weights can be taken as real and the dynamics remain sufficiently stable without additional bias-control techniques.
For realistic nuclear Hamiltonians, however, variance control and stability enhancements analogous to those used in FCIQMC---such as the initiator approximation and adapted shift method~\cite{FCIQMC_Hu}---can be incorporated into PTQMC in a straightforward manner. This provides a clear route toward applications in nuclear matter and other large-scale ab initio systems.

\bibliographystyle{modified-apsrev4-2}
\bibliography{reference_sm}